\documentclass[preprint]{acmart}
\AtBeginDocument{%
  \providecommand\BibTeX{{%
    \normalfont B\kern-0.5em{\scshape i\kern-0.25em b}\kern-0.8em\TeX}}}

\copyrightyear{2020} 
\acmYear{2020} 
\setcopyright{acmlicensed}\acmConference[ICT4S2020]{7th International Conference on ICT for Sustainability}{June 21--26, 2020}{Bristol, United Kingdom}
\acmBooktitle{7th International Conference on ICT for Sustainability (ICT4S2020), June 21--26, 2020, Bristol, United Kingdom}
\acmPrice{15.00}
\acmDOI{10.1145/3401335.3401359}
\acmISBN{978-1-4503-7595-5/20/06}

\begin{document}

\title{The United Nations Sustainable Development Goals in Systems Engineering: Eliciting sustainability requirements.}

\author{Ian Brooks}
\email{Ian.Brooks@uwe.ac.uk}
\orcid{0000-0002-6227-327X}
\affiliation{%
  \institution{University of the West of England}
  \streetaddress{Coldharbour Road}
  \city{Bristol}
\country{United Kingdom}
\postcode{BS36 1QG}
}
\renewcommand{\shortauthors}{Brooks.}
\begin{abstract}
This paper discusses a PhD research project testing the hypothesis that using the United Nations Sustainable Development Goals (SDG) \cite{UnitedNationsGeneralAssembly2015} as explicit inputs to drive the Software Requirements Engineering process will result in requirements with improved sustainability benefits.  The research has adopted the Design Science Research Method (DSRM) \cite{Wieringa2010} to test a process named SDG Assessment for Requirements Elicitation (SDGARE). Three DSRM cycles are being used to test the hypothesis in safety-critical, high-precision, software-intensive systems in aerospace and healthcare. Initial results from the first two DSRM cycles support the hypothesis. However, these cycles are in a plan-driven (waterfall) development context and future research agenda would be a similar application in an Agile development context.
\end{abstract}

\ccsdesc[500]{Software and its engineering~Requirements analysis}
\ccsdesc[500]{Social and professional topics~Sustainability}

\keywords{sustainability, requirements engineering, sustainable development goals, SDG, aerospace, cancer care}

\maketitle

\section{Doctoral Program}
The researcher is registered as a part-time PhD student at the University of the West of England, Bristol UK (UWE Bristol) in the Department of Computer Science and Creative Technologies (CSCT). The researcher is a Senior Lecturer in Sustainable IT and a member of the Software Engineering (SE) Research Group at UWE Bristol. Studies commenced in October 2016. Director of Studies is Prof James Longhurst (UWE Bristol) and the supervisory team includes Prof Mohammed Odeh (UWE Bristol) and Dr Mario Kossmann (AIRBUS).  The research fees are partly met by CSCT.  This research has not received funding from AIRBUS.

\section{Context and Motivation}
Software has an immense impact on our modern lifestyles.  It also has a significant impact on sustainability - economic, environmental and social \cite{Globale-SustainabilityInitiative2015} .  This impact can be positive e.g. carbon emissions reduction through digital delivery of music \cite{Weber2010} – or negative e.g. Volkswagen diesel emissions software cheating \cite{VolkswagenAktiengesellschaft2015}. The Smarter2030 report \cite{Globale-SustainabilityInitiative2015} concluded that Information and Communications Technology (ICT) will be required to achieve at least 20\% of the necessary global greenhouse gas (GHG) emissions reductions by 2030.  Most, perhaps all, of that ICT will have software at its heart.  

Software is unlikely to maximise delivery of sustainability improvements unless this is deliberately planned and specifically included in software requirements specifications.  We need to ensure that software engineers have SE methods which will support delivering positive sustainability benefits in the engineered software applications. There is a small body of research into approaches to embed or ‘infuse’ sustainability into requirements engineering (RE) \cite{Penzenstadler2014a}.

In 2015, the United Nations Sustainable Development Goals (SDGs) were adopted by 193 countries \cite{UnitedNationsGeneralAssembly2015}.  They consist of 17 goals and 169 targets describing the ‘World We Want’ by 2030.  Given that the SDGs could be described as a globally-adopted definition of sustainability, the researcher aimed to explore whether the SDGs could enhance the elicitation of sustainability requirements in SE.  If so, how could their use be effective in industry practice?

\section{Research Objectives}
The Research Questions being explored are as follows:

RQ1. Are the UN SDGs well-written as requirements for the engineering of sustainable software-intensive systems?

RQ2. Can the UN SDGs be semantically enriched and represented as a generic ontological model?

RQ3. What appropriate RE methods can be employed to develop sustainable software requirements specifications based on UN SDGs?

RQ4. Can the UN SDGs improve the sustainability benefits of software-intensive system requirements specifications using representative and sufficient enough case studies?

RQ5. What metrics can be adopted and generalised to assess the impact of using the UN SDGs in critically evaluating their role in developing sustainable software-intensive requirements specifications?

\section{Background}
\subsection{The Global Sustainability Crisis}
The world faces an existential sustainability crisis.  Steffen et al (2015) have documented the “safe operating space for humanity” and we have already exceeded the safe limits on three parameters relating to biosphere integrity (Genetic Diversity) and biogeochemical flows (nitrogen and phosphorus cycles) \cite{Steffen2015} whilst others including climate change are approaching the maximum safe limit. The scientific consensus on the climate change crisis is well documented in the reports of the Intergovernmental Panel on Climate Change, such as in their Assessment Report 5 \cite{IntergovernmentalPanelonClimateChange2014}.  

\subsection{United Nations Sustainable Development Goals: A framework for hope}
The SDGs, adopted in September 2015, represent a globally agreed set of sustainability aims with significant global political and business commitment \cite{UnitedNationsGeneralAssembly2015}.  All 193 countries which ratified the goals will be reporting on their progress.   

Although the SDGs themselves do not have the status of international law, inevitably they set the legal (soft law) and stakeholder expectations to 2030.  Indeed, some commentators have gone so far as to describe them as the “new global morality” \cite{Redo2015}.  From the SE perspective, the SDGs appear to be a great advance.  They are a global, comprehensive framework for sustainability.  If they can be used as software requirements, perhaps we can provide software engineers with the framework in which to deliver sustainability benefits in the course of their work.

In order to explore the hypothesis further, relevant literature has been reviewed to explore four core themes.

\subsection{SDGs and Requirements Engineering}
This is the core area of this PhD research and there is very little literature published in this area. A Scopus All Fields search on ("sustainable development goal") AND ("requirements engineering") returned only 31 papers (1 Feb 2020). Most of these appear because of a single mention of the SDGs in the text.  Only 13 of the papers had any realistic research significance. Two papers discussed SDGs as motivation for RE work, five mentioned the existence of the SDGs as a framework, six discussed the contribution ICT can make to delivering specific SDGs.  Disturbingly one paper mentioned SDGs in the abstract but not in the body of the paper, potentially a case of ‘greenwashing’.  None covered the use of the SDGs to elicit requirements. This indicates that this research project is addressing a gap in the knowledge base.

\subsection{SDGs and Software Engineering}
A similar search on the phrases “software engineering” and “sustainable development goal” published from 2014 onwards (the year before the SDGs were adopted) returned 139 citations (23 Feb 2020).  Again, most of these appear because of a single mention of the SDGs in the text.  A few papers do refer to the SDGs in the context of exploring generalised sustainability frameworks for SE such as Oyedeji, 2019 \cite{Oyedeji2019}.  Two of the papers were those published by the researcher. There were no papers reporting the direct use of the SDGs in SE. 

\subsection{Software Engineering for Sustainability (SE4S)}
SE4S is a more developed field as shown in the systematic review by Penzenstadler (2014) \cite{Penzenstadler2014b}.  Penzenstadler is one of the original authors of the Karlskrona Manifesto for Sustainability Design \cite{Becker2014} in software systems which sets out nine principles and commitments for the “group of people who design the software systems that run our world”.     “Requirements: The Key to Sustainability” by Becker et al (2016) \cite{Becker2016} further develops the SE4S ideas and alongside Nauman’s GREENSOFT model \cite{Naumann2011} these publications   form a foundation for this research project.

\subsection{SDGs and business impact}
In the five years since the adoption of the SDGs, a body of literature has started to be published.  The initial case study for this research is in the aerospace industry.  A Scopus search ( ALL ( "sustainable development goal" )  AND  TITLE-ABS-KEY ( aerospace )  OR  TITLE-ABS-KEY ( aviation ) )  returns just 36 citations.  Many of these address the conundrum of “sustainable tourism” and the current levels of climate change damage caused by aircraft.  These address specific targets within SDGs 12 Responsible Production and Consumption and 13 Climate Action.

The longer timescale of the academic publishing cycle on a specific initiative such as the SDGs, leaves a gap which is more rapidly filled by consultancies, trade bodies and activist groups.   These are making a useful contribution to the field, for example the report Flying in Formation – Air Transport and the SDGs \cite{AirTransportActionGroup2017} does review each of the 17 SDGs, identifying ideas for action, proposing a government agenda, giving examples of action in the sector and assessing the relative impact of the sector on the goal. 

Consultancies such as PwC are publishing, or contributing to, useful analysis such as in the reports The SDG Investment Case \cite{UNEPFinanceInitiative2017}  and Business Reporting on the SDGs \cite{GRI2017}.

\section{Hypothesis}
The hypothesis at the heart of this research is: 
Using the UN Sustainable Development Goals as explicit inputs to drive the Software RE process results in requirements with improved sustainability benefits.

\section{Research approach and methods}
\subsection{Design Science Research Method (DSRM)}
The core research approach is the use of SE case studies to demonstrate the difference between requirements developed under business as usual conditions and those informed by the SDGs.  Using the SDGs as an input to the RE process should result in a stronger, more holistic set of requirements for sustainability than the set developed without the SDGs. It is not expected that every SDG will be relevant in all cases. 

The methodology adopted is the Design Science Research Method (DSRM)  \cite{Wieringa2010}.  DSRM provides a recognised approach to testing designed artefacts, such as a requirements set, in this case before and after including the SDGs.  An All Fields SCOPUS search on “design science research method” returns 443 sources (23 Feb 2020).  
DSRM does have a notable weakness for this research in that it is explicitly human-centred and hence may not give sufficient weight to non-human ecosystem sustainability.

The RE method for this study is based upon that used by Kossmann in his doctoral thesis “OntoREM : An Ontology-Driven RE Methodology Applied in the Aerospace Industry” \cite{Kossmann2010}.  OntoREM is an approach which UWE Bristol has co-invented and for which UWE Bristol has a joint US patent with AIRBUS.   OntoREM was selected as it has been designed as a knowledge-driven approach to RE which has been shown to lead to “improved quality requirements in considerably less time and hence at less cost per requirement, than [...] a traditional RE process.” \cite[p ii]{Kossmann2010}. OntoREM is being used as a test scaffold. Future research should explore use with other RE methods.

The research programme consists of three DSRM cycles applying the SDGs as requirements in cases of safety-critical, high-precision, software-intensive systems. The artefact being designed, developed and tested has been named SDG Assessment for Requirements Elicitation (SDGARE). The SDGARE process is shown in figure 1.  The stages of SDGARE tested by each of the DSRM cycles is shown in figure 2. The SDGs themselves are not being redesigned though their strengths and weaknesses in practice will be reported.

\begin{figure}[ht]
  \centering
  \includegraphics[width=\linewidth]{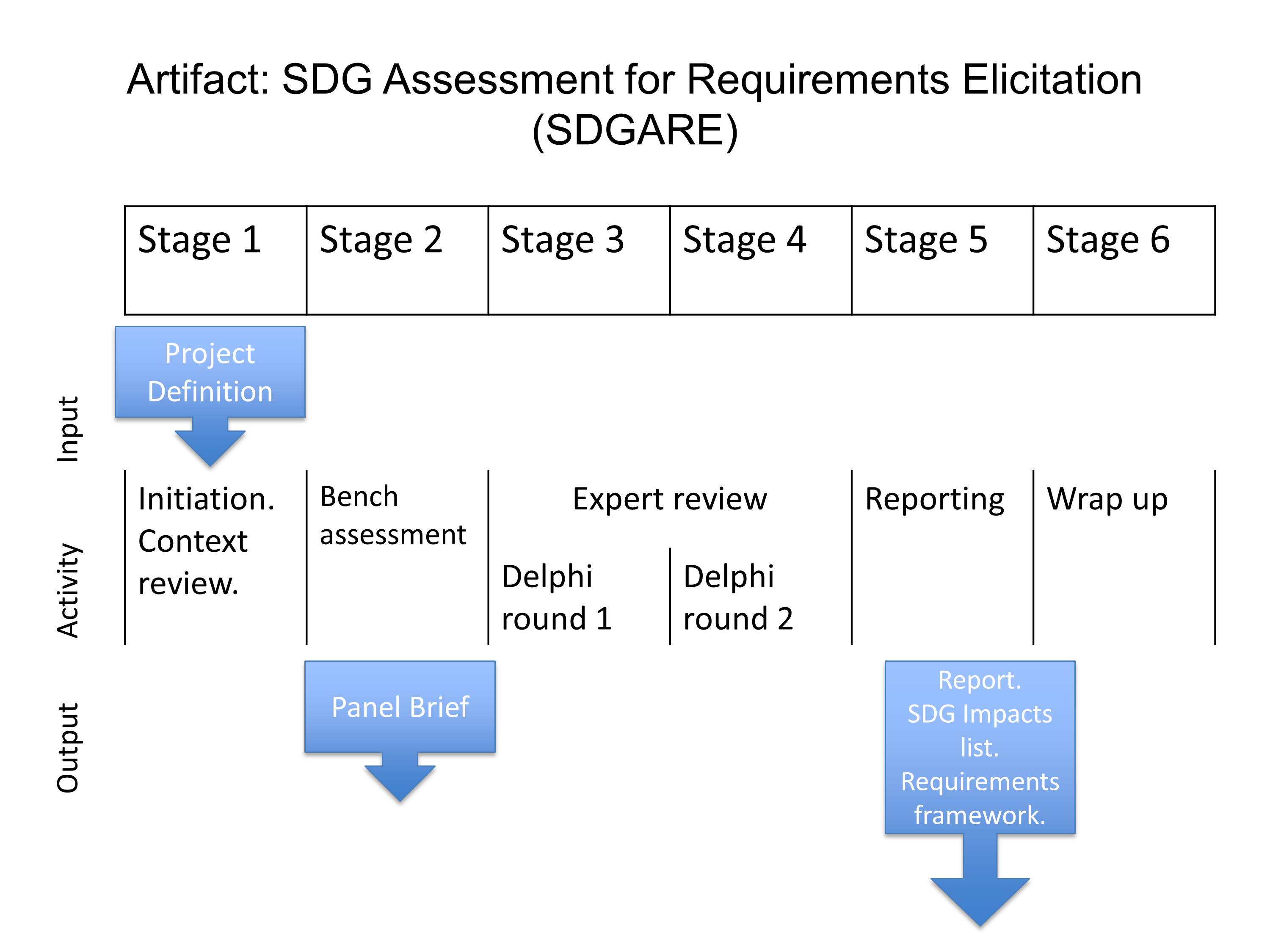}
  \caption{Stages of SDG Assessment for Requirements Elicitation (SDGARE).}
  \Description{Stages of SDG Assessment for Requirements Elicitation.}
\end{figure}

\begin{figure}[ht]
  \centering
  \includegraphics[width=\linewidth]{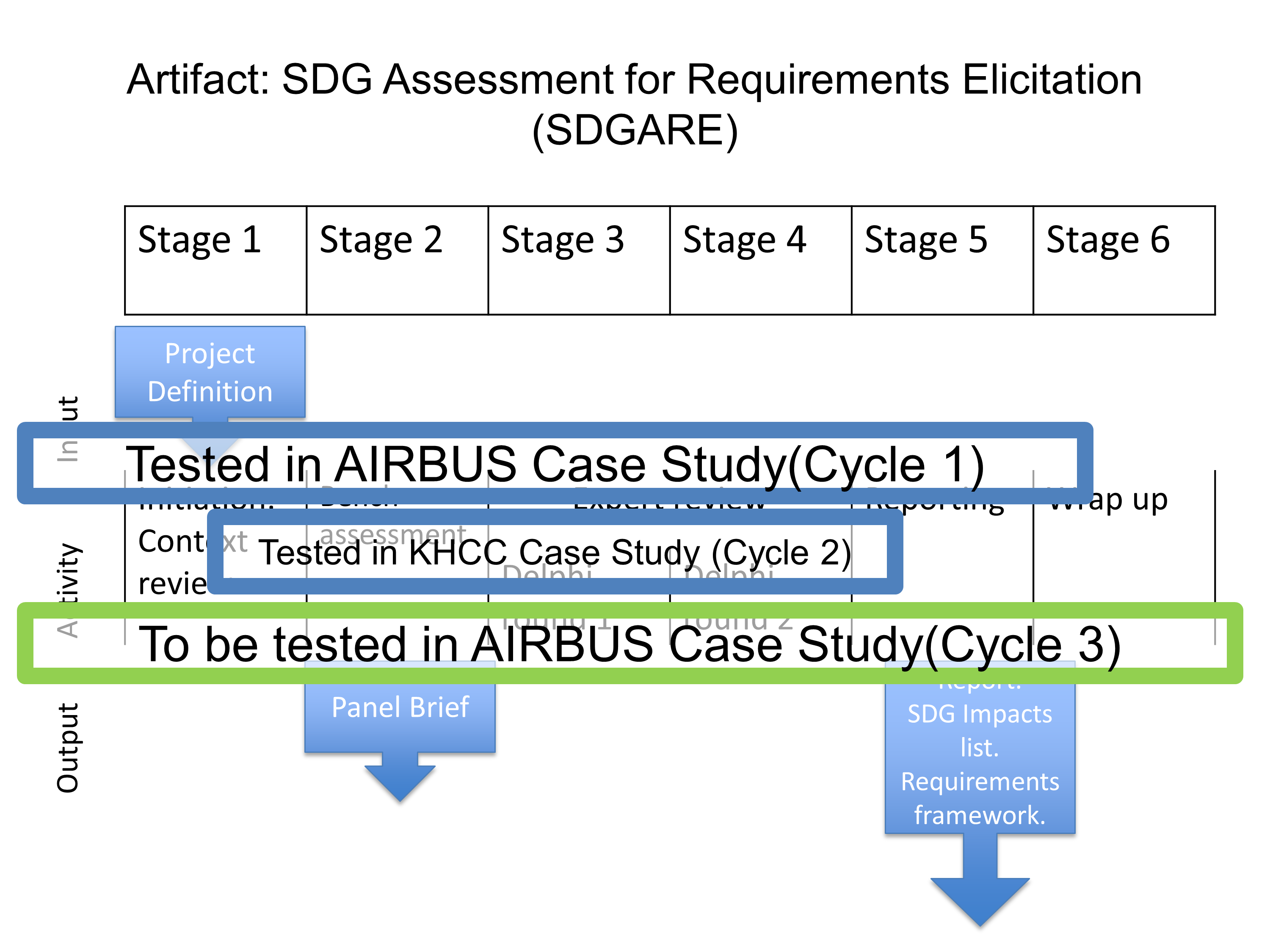}
  \caption{DSRM Cycles testing SDGARE.}
  \Description{DSRM Cycles testing SDGARE.}
\end{figure}

\subsection{Cycle 1 AIRBUS Additive Layer Manufacturing}
The first DSRM cycle tested an approach to using the SDGs to elicit requirements for the design and operation of a new Additive Layer Manufacturing (ALM) facility being planned by AIRBUS for their Filton site in Bristol, UK \cite{Airbus2019}. The setting was selected as matching the sought for case criteria.  AIRBUS were undertaking the requirements definition for the plant, using the OntoREM method (RQ3) and their business-as-usual stakeholder requirements elicitation.  The researcher ran a parallel process reviewing each of the 169 SDG targets for relevance to the ALM project.  The resulting requirements were merged with the requirements from the BAU process and put through the AIRBUS acceptance process.  The combined, accepted requirements set was analysed and coverage of the SDGs compared to evaluate whether the SDGARE process would result in improved sustainability benefit.  The SDG requirements were created as an ontological model through the OntoREM method (RQ2).

\subsection{Cycle 2 King Hussein Cancer Center, Cancer Care Informatics}
The second DSRM cycle tested an improved approach to SDGARE, adding a Delphi-style process for identifying the areas of SDG impact for a system.  In this case considering the SDG impacts of the Cancer Care Informatics (CCI) system in the context of King Hussein Cancer Center, Amman, Jordan. This setting also matched the sought for case criteria.  The Delphi panel consisted of both Cancer Care specialists and Sustainability experts and considered the impacts at SDG goal-level. The outcome was evaluated against a literature review of the CCI scope (RQ1, RQ5).

\subsection{Cycle 3 AIRBUS ALM}
The third and final DSRM cycle will be a second run of the AIRBUS ALM case to evaluate the effectiveness of SDGARE when using a Delphi process operating at the SDG target-level.  The updated approach will be evaluated by comparing the resulting requirements against the output of the first cycle and by acceptability of the requirements to the AIRBUS ALM project leader.

\section{Results to date}
\subsection{Cycle 1 AIRBUS ALM}
The initial version of SDGARE was applied in parallel with the BAU requirements elicitation process as described above.  The BAU process elicited requirements covering four of the SDGs.  The SDGARE process elicited requirements covering these four and an additional six SDGs (RQ5).  This finding supports the hypothesis (RQ4). However, this DSRM cycle also identified that the SDGs are not all well-formed as requirements and that turning the global-level SDG targets into project-specific, quantified requirements was problematic (RQ1). The SDGs are far from a simple checklist when applied at project level. This version of SDGARE relied on the knowledge and experience of the researcher and would be difficult to replicate at scale.  The issues of translating SDG targets into requirements was presented at the INCOSE 2018 conference \cite{Brooks2018}.

\subsection{Cycle 2 KHCC CCI}
This DSRM cycle experimented with the use of a Delphi-style approach to overcome the limitation of the scarce pool of individuals with the experience to apply the initial design of SDGARE.  This process successfully identified a wide range of impacts of, and hence requirements on, CCI systems (RQ4).  The process enables more interchangeability of people and expertise and so could more easily be scaled up.  

\subsection{Validity of results}
Clearly these two DSRM cycles are small in scale and it is fair to question how representative the results would be of widescale application of SDGARE.  However, AIRBUS has a culture and business management system that emphasises regulatory compliance and the BAU process would be expected to provide good coverage of core SDG targets, as shown.  That the SDGARE process led to enhanced coverage of the SDGs, even in this compliance-oriented culture, provides reason to believe that improved sustainability requirements could be elicited through SDGARE in most organisations.
The AIRBUS cycles are in the context of a plan-based (waterfall) development approach.  The question of the contribution of the SDGs in an Agile development approach should be a future research agenda. Future research should also validate the approach in other industries and other RE methods.

\section{Dissertation status and next steps}
The researcher has passed the first and second progress examinations required by the UWE Bristol PhD research programme (2018 and 2019). The next step, to undertake DSRM Cycle 3 applying a Delphi-based approach at SDG target level on the AIRBUS ALM case study, is expected to be completed by the end of 2020.  The researcher will then focus on writing up the thesis with a target completion of Summer 2021.

\section{Current and expected contributions}
Current contributions have included the peer-reviewed conference papers \cite{Brooks2018} and \cite{Brooks2018a}. Future contributions will include a journal paper on the third DSRM cycle using a Delphi process with the SDGs in the context of AIRBUS’ ALM facility development.

\bibliographystyle{ACM-Reference-Format}
\bibliography{ICT4S-2020-DS-refs}

\end{document}